\documentclass[conference]{IEEEtran}
\IEEEoverridecommandlockouts
\usepackage{cite}
\usepackage{amsmath,amssymb,amsfonts}
\usepackage{algorithmic}
\usepackage{graphicx}
\usepackage{textcomp}
\usepackage{xcolor}
\usepackage{listings}
\usepackage{amsmath}
\usepackage{algorithmic}
\usepackage{graphicx}
\usepackage{textcomp}
\usepackage{xcolor}
\usepackage{mathtools}
\usepackage{xspace}
\usepackage{float}
\usepackage{subfig}
\usepackage{makecell}
\usepackage{lipsum}
\usepackage{soul}
\def\BibTeX{{\rm B\kern-.05em{\sc i\kern-.025em b}\kern-.08em
    T\kern-.1667em\lower.7ex\hbox{E}\kern-.125emX}}
\makeatletter
\newcommand{\linebreakand}{%
  \end{@IEEEauthorhalign}
  \hfill\mbox{}\par
  \mbox{}\hfill\begin{@IEEEauthorhalign}
}
\makeatother
    
\newcommand{\reg}{$^{\circledR}$}
\newcommand{\spec}{SPEC CPU\reg\ 2017\ }
\newcommand{\specspeed}{SPECspeed\reg\ 2017\ }
\newcommand{\specrate}{SPECrate\reg\ 2017\ }

\begin{document}

\title{The AMD Rome Memory Barrier}

\author{\IEEEauthorblockN{Phillip Allen Lane}
\IEEEauthorblockA{\textit{Department of Computer Science} \\
\textit{The University of Alabama in Huntsville}\\
Huntsville, Alabama, USA \\
phillip.lane@uah.edu}
\and
\IEEEauthorblockN{Jessica Lobrano}
\IEEEauthorblockA{\textit{Department of Electrical and Computer Engineering} \\
\textit{The University of Alabama in Huntsville}\\
Huntsville, Alabama, USA \\
jal0040@uah.edu}
}

\maketitle

\begin{abstract}
With the rapid growth of AMD as a competitor in the CPU industry, it is imperative that high-performance and architectural engineers analyze new AMD CPUs. By understanding new and unfamiliar architectures, engineers are able to adapt their algorithms to fully utilize new hardware. Furthermore, engineers are able to anticipate the limitations of an architecture and determine when an alternate platform is desirable for a particular workload. This paper presents results which show that the AMD ``Rome'' architecture performance suffers once an application's memory bandwidth exceeds 37.5 GiB/s for integer-heavy applications, or 100 GiB/s for floating-point-heavy workloads. Strong positive correlations between memory bandwidth and CPI are presented, as well as strong positive correlations between increased memory load and time-to-completion of benchmarks from the \spec benchmark suites.
\end{abstract}


\section{Introduction}

In the last five years, AMD has made immense strides in the CPU industry. As recent as 2016, AMD had little to no competitive products in the desktop and server CPU market. Their most powerful options were the architectures known as Piledriver, Bulldozer, Excavator, and Steamroller. These offered profoundly poor IPC compared to Intel offerings at the time. These architectures promised high core-counts; in reality, pairs of 2 cores called ``compute units'' shared instruction cache and fetch hardware, and all of these except Steamroller shared decode hardware within a compute unit~\cite{uarch}. This resulted in processors much closer to a lower core count processors with simultaneous multi-threading (SMT). These architectures were especially power-inefficient; for instance, the 8-thread FX-9590 CPU had a thermal design power of 220 watts while failing to perform on par with an Intel Core i7 from the same era that emitted less than half the heat~\cite{fx}.

With the release of the Zen architecture in February 2017, Zen delivered a high-performance and competitive architecture with higher core counts than their competition. In August 2019, the second generation of AMD Epyc released, featuring up to 64 cores and 128 threads on a single socket. Known as Rome, second-generation Epyc provided significant architectural improvements over first-generation Naples processors.

As novel microarchitectures are released, there are many associated difficulties. Researchers and engineers must rapidly adapt their software to make optimal use of new hardware. Writers of certain pieces of software such as compilers and math libraries are especially affected. Widespread use of these types of software means that extreme optimization and efficient use of the underlying hardware is critical. However, the first step of tuning code for a specific architecture is to first gather characteristic data about the architecture with the aim that analysis may be performed.

This paper aims to provide support for the ``memory barrier'' of AMD Rome, i.e. the maximum memory bandwidth that can be sustained without extreme penalties to the IPC of an application. The \spec benchmarks were compiled and run on a second-generation AMD Epyc system, and results were analyzed. Section~\ref{bg} provides some insight into prior work and background about the Rome microarchitecture. Section~\ref{spec} discusses the \spec benchmarks used as the basis for performance measurement in this work. Section~\ref{setup} provides further details on the particular system used for testing, as well as testing methodology. Section~\ref{results} discusses the results from the testing. Finally, section~\ref{conclusions} provides conclusions and potential directions for future work.

\section{Background} \label{bg}
\subsection{SPEC Analysis on Intel i7-8700k}
Hebbar and Milenkovi\'{c} performed SPEC\reg\ analysis on an Intel Core i7-8700k "Coffee Lake" processor~\cite{ranjan}. Their work demonstrated that the Intel compilers were significantly faster than their GNU counterparts on an Intel system. Their research illustrated system bottlenecks in each benchmark from SPEC CPU\reg\ 2017, such as whether an application was memory-bound, front-end bound, or bound elsewhere. In their results, they found the top memory-bound benchmarks included 607.cactuBSSN\_s, 619.lbm\_s, 649.fotonik3d\_s, 654.roms\_s, 620.omnetpp\_s, 623.xalancbmk\_s, and 657.xz\_s.

\subsection{Memory Characterization of SPEC CPU2017}
Singh and Awasthi~\cite{specmem} performed a deep analysis of the \spec benchmark memory characteristics through a mix of dynamic binary instrumentation, performance counter monitoring, and OS-based monitoring. Their findings showed \specspeed workloads had significantly more instructions than their \specrate counterparts, and that floating-point benchmarks were more efficient IPC-wise than the integer benchmarks. They also showed benchmarks with a large working set size (WSS) and high memory requirements have lower instructions per clock (IPC). They also showed the \specspeed benchmarks had a much higher WSS than the \specrate benchmarks. They further demonstrated the benchmarks 603.bwaves\_s, 607.cactuBSSN\_s, 649.fotonik3d\_s, 654.roms\_s, and 657.xz\_s had large memory footprints. Finally, they showed that 605.mcf\_s, 607.cactuBSSN\_r, 649.fotonik3d\_s, and 657.xz\_s, had heavy off-chip traffic and could thus be used to analyze memory bandwidth performance.

\subsection{Rome Cache Hierarchy}
\begin{figure}[tbh]
    \centering
    \includegraphics[width=.48\textwidth]{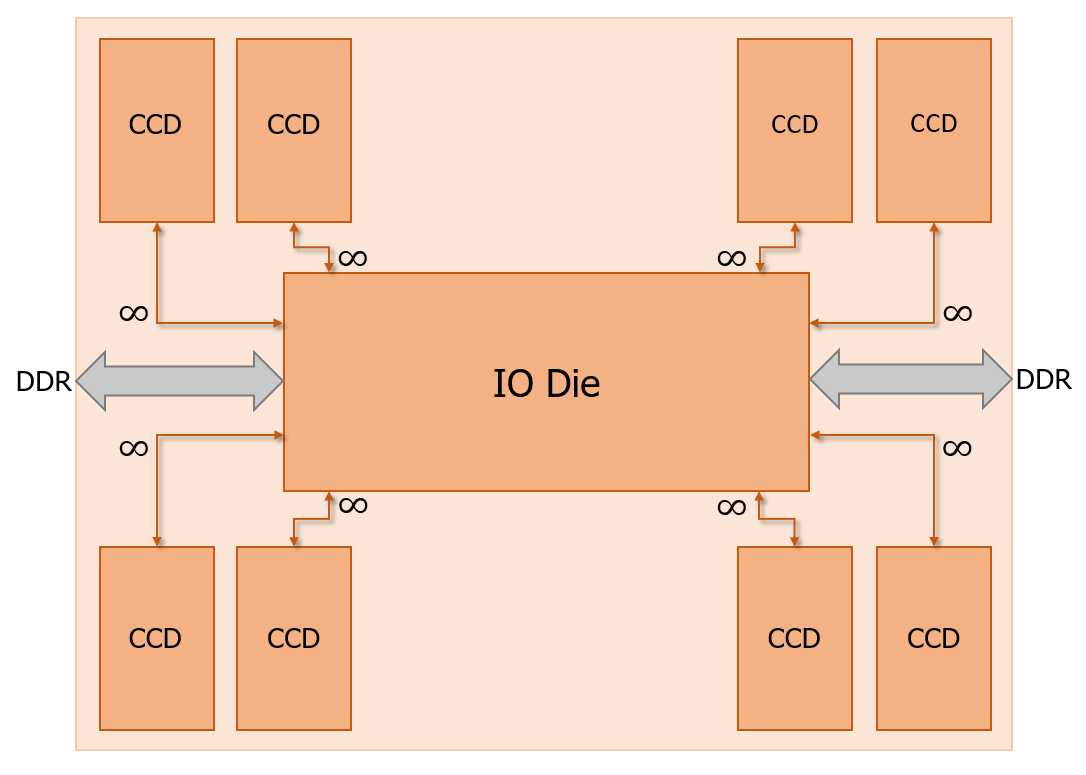}
    \caption{Diagram showing CCD hierarchy and communication through the IO die. The $\infty$ symbol represents AMD's Infinity Fabric interconnect.}
    \label{fig:milanpack}
\end{figure}

The second-generation Epyc processors have cache hierarchies which are especially notable for exposing extreme NUMA (Non-Uniform Memory Access) phenomena. The top-end Epyc 7742 has 9 physical dies per package, 8 for cores and 1 for IO~\cite{7742}. Each of the 8 core dies, called a Core Complex Die (CCD), contains 8 cores. Within a CCD exist two CPU Complexes (CCXs), which contain 4 cores. Each CCX shares a physical L3 cache, and it is important to note that despite two CCXs sharing a die, they do not have direct access to each other's cache and must route through the IO die~\cite{romecache}. Logically, the entirety of the L3 cache on the processor is shared, but caches on separate dies must also route through the IO die to share memory~\cite{zen2}. This manifests as a complicated NUMA chip that is difficult to predict. Figure~\ref{fig:milanpack} shows how the CCDs communicate with each other and with main memory through the IO die.

As for the cache hierarchy itself, each core has private L1I, L1D, and L2 caches. The L1I and L1D caches are 32 KiB 8-way set associative caches, with the L1D being write-back. The L2 cache is a unified 512 KiB 8-way set associative cache, also being write-back. The L3 is 16 MiB per CCX, yielding a total of 128 MiB total L3 cache. However, we emphasize this 128 MiB of L3 cache is segmented across 8 different CCXs. Finally, the cross-socket communication technologies of AMD fall short of Intel’s technologies, which means many benchmarks will struggle to scale efficiently across both sockets of dual-socket machines.

The ``chiplet'' design that AMD uses incorporates the use of multiple dies per package, which makes high core count CPUs significantly cheaper and easier to manufacture. For instance, the 48-core AMD Epyc 7643 is approximately \$5000~\cite{7643}, while the 40-core Intel Xeon Platinum 8380 is approximately \$8100~\cite{8380}. This price difference increases as the core counts rise. However, the downsides of the chiplet design is the manifestation of significantly higher latency core-to-core communication and complex NUMA phenomena arising from the segmented L3 caches, both of which cannot be ignored.

\subsection{Prior Analysis on Rome Cache Hierarchy}
Velten, et al. demonstrated that the Rome cache hierarchy displays extreme NUMA effects~\cite{romecache}. In their work, the explained that AMD uses a MDOEFSI (Modified, Dirty, Owned, Exclusive, Forward, Shared, Invalid) cache coherence protocol. Access to locally-owned L1, L2, and L3 caches take 4, 12, and 39 cycles respectively. However, access to a remote L3, which is often necessary due to Rome's distributed L3 architecture, takes a minimum of 200 clock cycles, and is often on par with simply reading from main memory. In some cases where access to a cache line is in a modified or exclusive state, access to a remote L3 can exceed the latency of reading from main memory.

However, for local cache accesses, latencies are often lower than Intel Cascade Lake-X (CLX), though accesses to main memory are about 20 clock cycles slower, and is suspected to result from the requirement to route all memory accesses through the IO die over the Infinity Fabric. Velten, et al. propose Rome as a preferred architecture for applications with high memory demands, insofar as inter-core communication is kept to a minimum.

\subsection{S-UMA Alternative to AMD's Solution}
Figure~\ref{fig:ccdarch} shows the cache hierarchy inside a CCD in a Rome-based AMD Epyc 7402. Fotouhi, et al. propose an alternative solution which includes integrated silicon-photonic interconnects and migration of shared L3 cache to a separate die (S-UMA) to overcome many of AMD's limitations~\cite{chipletcache}. By using a low-latency solution with distance-independent energy consumption, their solution in simulations was able to achieve an average speedup of 12\% compared to AMD Rome, and up to 30\% power savings on an 8-die 64-thread processors. The existence of prior literature demonstrating alternatives to Rome's solution demonstrates the inherent limitations of AMD's current solution with a distributed L3 cache.

\subsection{Choosing a Compiler}
Halbiniak, et al. demonstrated the performance impact by choice of compiler on an AMD Epyc 7742 (Rome) processor~\cite{compilers}. Their work proved that the Intel \texttt{icpc} compiler outperforms all other compilers tested, and better utilizes the AVX2 instruction set extension on Rome as opposed to the AMD \texttt{aocc} compiler. The Intel compiler performed $1.3\times$ better than the AMD compiler on AMD's own architecture when compiling solidification numerical modeling code.

\section{SPEC CPU2017} \label{spec}
Benchmarks have been used to evaluate and measure performance comparisons across different computer architectures. From~\cite{ranjan}~\cite{spec}, the Standardized Performance Evaluation Corporation (SPEC\reg) works to create representative benchmarks of common high-performance computing (HPC) applications to evaluate the performance of CPUs. Founded in 1988, the non-profit organization aims to “establish, maintain and endorse standardized benchmarks and tools to evaluate performance and energy efficiency” for computer systems [2]. Over the years, there have been six published releases of the SPEC CPU\reg\  benchmark suites: CPU\reg 89, CPU\reg 92, CPU\reg 95, CPU\reg 2000, CPU\reg 2006, and CPU\reg 2017. Each new release provided more complexity and workloads designed to address the advances in both software and hardware for computer systems. Derived from real applications the benchmarks provide the standard for uniform CPU intensive workloads to measure and compare different systems. 

Many ways exist to measure a computer system’s performance, two of the most common ways are time (seconds to complete a workload) and throughput (work completed per unit time).  CPU\reg 2017 contains 43 individual benchmarks which are organized into four sub-suites which focus on different types of computer intensive performance: floating point rate, floating point speed, integer rate, and integer speed benchmarks. Table~\ref{tab:spec} below illustrates the various benchmark suites.

\begin{table*}
\small
\centering
\caption{List of \spec benchmark suites.}
\label{tab:spec}
\begin{tabular}{|l|l|l|l|}\hline
\textbf{Suite} & \textbf{Contents}      & \textbf{Metrics}      & \textbf{Notes}    \\ \hline
\makecell{\specspeed \\ Integer} & 10 integer benchmarks & \makecell{ SPECspeed\reg 2017\_int\_base \\ SPECspeed\reg 2017\_int\_peak \\ SPECspeed\reg 2017\_int\_energy\_base \\ SPECspeed\reg 2017\_int\_energy\_peak } & \makecell{ The \specspeed suites run \\ one copy of each benchmark. \\ Higher scores indicate less \\ execution time is required. }\\ \hline
\makecell{\specspeed \\ Floating Point} & 10 floating point benchmarks & \makecell{ SPECspeed\reg 2017\_fp\_base \\ SPECspeed\reg 2017\_fp\_peak \\ SPECspeed\reg 2017\_fp\_energy\_base \\ SPECspeed\reg 2017\_fp\_energy\_peak } & \makecell{ The \specspeed suites run \\ one copy of each benchmark. \\ Higher scores indicate less \\ execution time is required. }\\ \hline
\makecell{\specrate \\ Integer} & 10 integer benchmarks & \makecell{ SPECrate\reg 2017\_int\_base \\ SPECrate\reg 2017\_int\_peak \\ SPECrate\reg 2017\_int\_energy\_base \\ SPECrate\reg 2017\_int\_energy\_peak } & \makecell{ \specrate suites run  \\ multiple concurrent copies \\ of each benchmark. Higher scores \\ indicate increased throughput (more work \\ done per unit of time). }\\ \hline
\makecell{\specrate \\ Floating Point} & 13 floating point benchmarks & \makecell{ SPECrate\reg 2017\_fp\_base \\ SPECrate\reg 2017\_fp\_peak \\ SPECrate\reg 2017 2017\_fp\_energy\_base \\ SPECrate\reg 2017\_fp\_energy\_peak } & \makecell{ \specrate suites run  \\ multiple concurrent copies \\ of each benchmark. Higher scores \\ indicate increased throughput (more work \\ done per unit of time). }\\
\hline
\end{tabular}
\end{table*}

Both \specspeed and \specrate benchmark ratios are averaged using the geometric mean and reported as the overall metric for the given benchmark. The rate benchmarks are designed to stress the throughput of the application data type, while the speed benchmarks stress the speed of each application type. The base metrics require benchmarks in any given suite of any given language be compiled with the same flags in the same order. The optional peak metrics allow for various compiler options to be used for each benchmark.

The \spec Integer and Floating Point benchmarks provide a wide range of different application areas ranging from a Perl interpreter to explosion modeling. These benchmarks are composed of various programming languages to provide a range of compiler optimization options. To break down the suites further, Table~\ref{tab:specint} below shows the breakdown of each of the integer suites’ benchmark details and Table~\ref{tab:specfp} below shows the breakdown of each of the floating point suites’ benchmark details.

\begin{table*}
\small
\centering
\caption{List of \spec integer benchmark suites.}
\label{tab:specint}
\begin{tabular}{|l|l|l|r|l|}\hline
\textbf{\makecell{\specrate \\ Integer}} & \textbf{\makecell{\specspeed \\ Integer}} & \textbf{Language} & \textbf{KLOC*} & \textbf{Application Area}\\ \hline
500.perlbench\_r & 600.perlbench\_s & C & 362 & Perl interpreter\\ \hline
502.gcc\_r & 602.gcc\_s & C & 1,304 & GNU C compiler \\ \hline
505.mcf\_r & 605.mcf\_s & C & 3 & Route planning \\ \hline
520.omnetpp\_r & 620.omnetpp\_s & C++ & 134 & Discrete Event simulation - computer network\\ \hline
523.xalancbmk\_r & 623.xalancbmk\_s & C++ & 520 & XML to HTML conversion via XSLT\\ \hline
525.x264\_r & 625.x264\_s & C & 96 & Video compression\\ \hline
531.deepsjeng\_r & 631.deepjeng\_s & C++ & 10 & Artificial Intelligence: alpha-beta tree search (Chess)\\ \hline
541.leela\_r & 641.leela\_s & C++ & 21 & Artificial Intelligence: Monte Carlo tree search (Go)\\ \hline
548.exchange\_r & 648.exchange2\_s & Fortran & 1 & Artificial Intelligence: recursive solution generator (Sudoku)\\ \hline
557.xz\_r & 657.xz\_s & C & 33 & General data compression\\
\hline
\end{tabular}
\end{table*}

\begin{table*}
\small
\centering
\caption{List of \spec floating-point benchmark suites.}
\label{tab:specfp}
\begin{tabular}{|l|l|l|r|l|}\hline
\textbf{\makecell{\specrate \\ Floating Point}} & \textbf{\makecell{\specspeed \\ Floating Point}} & \textbf{Language} & \textbf{KLOC*} & \textbf{Application Area}\\ \hline
503.bwaves\_r & 603.bwaves\_s & Fortran & 1 & Explosion modeling\\ \hline
507.cactuBSSN\_r & 607.cactuBSSN\_s & C++, C, Fortran & 257 & Physics: relativity\\ \hline
508.namd\_r & & C++ & 8 & Molecular dynamics\\ \hline
510.parest\_r & & C++ & 427 & Biomedical imaging: optical tomography with finite elements\\ \hline
511.povray\_r & & C++, C & 170 & Ray tracing\\ \hline
519.lbm\_r & 619.lbm\_s & C & 1 & Fluid dynamics\\ \hline
521.wrf\_r & 621.wrf\_s & Fortran, C & 991 & Weather forecasting\\ \hline
526.blender\_r & & C++, C & 1,577 & 3D rendering and amimation\\ \hline
527.cam4\_r & 627.cam4\_s & Fortran, C & 407 & Atmosphere modeling\\ \hline
& 628.pop2\_s & Fortran, C & 338 & Wide-scale ocean modeling (climate level)\\ \hline
538.imagick\_r & 638.imagick\_s & C & 259 & Image manipulation\\ \hline
544.nab\_r & 644.nab\_s & C & 24 & Molecular dynamics\\ \hline
549.fotonik3d\_r & 649.fotonik3d\_s & Fortran & 14 & Computational Electromagnetics\\ \hline
554.roms\_r & 654.roms\_s & Fortan & 210 & Regional ocean modeling\\
\hline
\end{tabular}
\end{table*}

\section{Test Setup} \label{setup}
\subsection{Euler}

\begin{figure}[tbh]
    \centering
    \includegraphics[width=.48\textwidth]{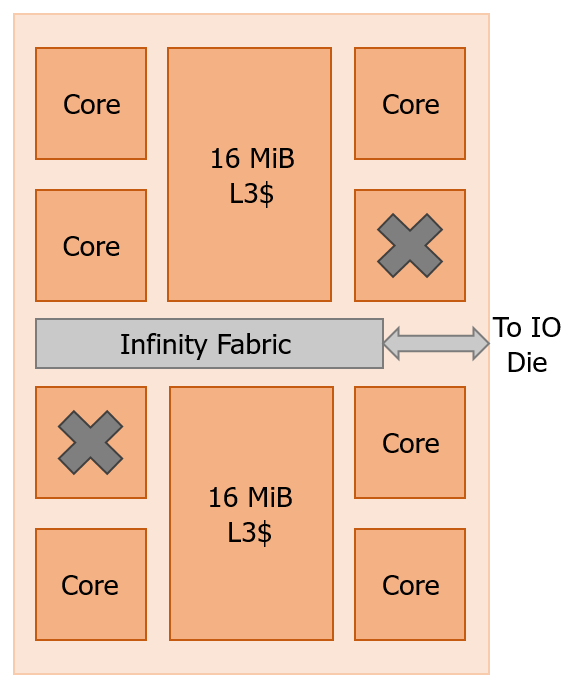}
    \caption{CCD architecture of AMD Epyc 7402 processor.}
    \label{fig:ccdarch}
\end{figure}

For our testing we used the University of Alabama in Huntsville's Electrical and Computer Engineering Department's \textit{Euler} machine. Euler is configured with two AMD Epyc 7402 processors and 256 GiB of memory. As a lower-tier Rome CPU, the Epyc 7402 has only four CCDs, and each CCD contains only 6 cores, or 3 cores per CCX. This yields a 24-core CPU with 128 MiB of L3 cache segmented across 8 CCXs. The dies used in an Epyc 7402 processor are binned dies due to one or two cores failing verification. Euler has two of these CPUs, so the 256 GiB of main memory is segmented into two NUMA (non-uniform memory access) regions, with the additional NUMA phenomena of the segmented L3 cache\cite{7402}. Therefore, Euler displays extreme NUMA phenomena which will prove difficult for engineers to write optimized code for. Figure~\ref{fig:ccdarch} shows the CCD architecture of the Epyc 7402 processor.

It is important to note, since Euler has 3 cores per L3 segment, its CPUs have a higher cache-per-core ratio than the higher-end 32 or 64 core models of AMD CPUs. For instance, the Epyc 7402 processor has about 5,460 KiB/core of L3 cache, while the Epyc 7502 with 32 cores has 4,096 KiB/core of L3 cache. Therefore, benchmarks that favor a higher cache-to-core ratio as opposed to higher core counts may prefer the Epyc 7402 as opposed to the Epyc 7502.

Euler is configured with a $16\times 16$ GiB memory configuration in 16 memory channels (8 per socket). Each module of RAM is a registered, dual-rank module capable of 3,200 MT/s. With $8\ \textrm{B/T}\times 3,200\ \textrm{MT/s}=25\ \textrm{GiB/s}$ memory transfer speed per channel, Euler is capable of up to $400\ \textrm{GiB/s}$ of memory bandwidth across both sockets. Because each socket is its own NUMA region, there are overheads associated with accessing a page on a remote NUMA region, so this peak memory bandwidth is rarely achieved in heavily threaded workloads.

Finally, The memory bandwidth counters on Euler are measured as total traffic across the data fabric (or Infinity Fabric). Thus, when we say ``memory bandwidth'' we refer to all accesses to both main memory and remote caches on a different CCX.

\subsection{AMD uProf}

AMD uProf is a tool for performance analysis and monitoring on AMD systems. AMD uProf allows engineers and researchers to profile an application for hotspot measurements, performance counters, and other characteristic data of an application. This allows engineers and researchers to optimize and tune their application to fit time and efficiency requirements. Alternatively, it can be used with a known benchmark suite to analyze the CPU's behavior and identify potential inefficiencies in a CPU's architecture.

We use a specific tool called AMDuProfPcm which allows more fine-tuned monitoring than the AMDuProfCLI ``collect'' tool used by~\cite{ranjan}. An XML file containing the events to be monitored is provided to the AMDuProfPcm tool, and the tool outputs a CSV of events and common metrics. AMDuProfPcm outputs core-specific metrics (including instructions retired, load/store operations, L2 cache access statistics, etc.), CCX-specific metrics (including L3 cache statistics), and statistics related to specific memory channels (including memory read-write bandwidth to a channel of memory).

\subsection{Test Methodology}

Two specific types of tests are performed--a scalability test  and monitoring test. The scalability test ran the rate benchmarks at several different instance counts $i$ where $i\in\{1,3,6,24,48\}$. Processes were pinned to cores in ascending order, so for $i=6$, the processes were pinned to CCD 0. For $i=1$, the test was designed to measure optimal conditions where one instance of the benchmark ran on one core with the rest of the system unstressed. For $i=3$, the test was designed to measure contention in a single CCX. For $i=6$, the test was similar to $i=3$ except we measured contention in a single CCD. For $i=24$, the test was designed to measure one complete package loaded, and $i=48$ stressed the entire system.

The monitoring test ran each of the SPEC ``base'' benchmarks under AMDuProfPcm. The SPEC rate benchmarks were run with 48 instances and all cores were monitored. The SPEC speed benchmarks were run with 1 instance with the number of threads set to 48. Because some SPEC speed benchmarks are single-threaded, not all speed monitoring tests stressed the entire system. For these cases, we opted not to run 48 instances of the single-threaded speed benchmarks because many of these benchmarks consume much more memory than the rate benchmarks. Additionally, we found that running 48 instances of certain speed benchmarks resulted in Euler running out of memory. Thus, for consistency, we used 1 instance for all speed benchmarks, regardless of whether or not the benchmark is threaded.

Due to difficulties associated with the AMDuProfPcm tool, benchmark 627.cam4\_s is not shown in the results. We expected that the AMDuProf ``collect'' tool would work, but we opted instead to omit this benchmark from our results because the ``collect'' tool is insufficiently powerful to collect the counters measured.

\subsection{Compilation of SPEC CPU2017}
\spec was compiled with the AOCC 3.1.0 compiler with the \texttt{-g -O3 -ffast-math -march=native -flto} flags for C/C++ sources and with the \texttt{-g -O3 -march=native -flto -Kieee -fno-finite-math-only} flags for Fortran sources. We compiled all speed benchmarks with parallelization enabled (if supported) and the necessary portability flags for our system. For the integer benchmarks (for both speed and rate) we added the \texttt{-fgnu89-inline -z muldefs} compile flags in addition to those aforementioned.

\section{Results} \label{results}
\subsection{Monitoring Test}

\begin{table*}
\centering
\caption{\spec floating-point rate cache results.}
\label{tab:specfprcache}
\begin{tabular}{|l|r|r|r|r|r|r|r|r|}\hline
\textbf{BM} & \textbf{L2 Access} & \textbf{L2 Miss} & \textbf{L2 Miss Rate} & \textbf{L2 Hit Rate} & \textbf{L3 Access} & \textbf{L3 Miss} & \textbf{L3 Miss Rate} & \textbf{L3 Hit Rate}\\ \hline
503 & 124.94 & 51.71 & 41.39 & 58.61 & 58.44 & 55.39 & 94.79 & 5.21\\ \hline
507 & 233.60 & 20.64 & 8.83 & 91.17 & 23.41 & 15.67 & 66.95 & 33.05\\ \hline
508 & 37.50 & 0.93 & 2.48 & 97.52 & 0.93 & 0.51 & 55.31 & 44.69\\ \hline
510 & 82.17 & 32.86 & 40.00 & 60.00 & 36.94 & 4.36 & 11.81 & 88.19\\ \hline
511 & 37.49 & 0.06 & 0.16 & 99.84 & 0.06 & 0.00 & 4.11 & 95.89\\ \hline
519 & 103.26 & 30.33 & 29.38 & 70.62 & 32.19 & 23.98 & 74.52 & 25.48\\ \hline
521 & 43.29 & 15.76 & 36.39 & 63.61 & 18.57 & 7.84 & 42.24 & 57.76\\ \hline
526 & 27.42 & 8.34 & 30.41 & 69.59 & 8.06 & 2.00 & 24.88 & 75.12\\ \hline
527 & 39.33 & 11.03 & 28.04 & 71.96 & 11.06 & 1.77 & 16.03 & 83.97\\ \hline
538 & 13.69 & 0.41 & 2.97 & 97.03 & 0.35 & 0.01 & 2.29 & 97.71\\ \hline
544 & 18.17 & 1.28 & 7.02 & 92.98 & 1.41 & 0.68 & 48.67 & 51.33\\ \hline
549 & 86.02 & 39.80 & 46.27 & 53.73 & 44.85 & 37.04 & 82.59 & 17.41\\ \hline
554 & 138.90 & 62.76 & 45.18 & 54.82 & 70.97 & 29.88 & 42.11 & 57.89\\ 
\hline
\end{tabular}
\end{table*}

\begin{table*}
\centering
\caption{\spec floating-point speed cache results.}
\label{tab:specfpscache}
\begin{tabular}{|l|r|r|r|r|r|r|r|r|}\hline
\textbf{BM} & \textbf{L2 Access} & \textbf{L2 Miss} & \textbf{L2 Miss Rate} & \textbf{L2 Hit Rate} & \textbf{L3 Access} & \textbf{L3 Miss} & \textbf{L3 Miss Rate} & \textbf{L3 Hit Rate}\\ \hline
603 & 49.53 & 21.08 & 42.55 & 57.45 & 22.64 & 21.31 & 94.13 & 5.87\\ \hline
607 & 160.65 & 8.42 & 5.24 & 94.76 & 8.47 & 5.95 & 70.29 & 29.71\\ \hline
619 & 113.84 & 36.08 & 31.69 & 68.31 & 36.26 & 32.84 & 90.58 & 9.42\\ \hline
621 & 30.32 & 10.84 & 35.76 & 64.24 & 12.69 & 4.45 & 35.64 & 64.36\\ \hline
628 & 33.06 & 10.16 & 30.73 & 69.27 & 11.32 & 2.84 & 24.92 & 75.08\\ \hline
638 & 29.80 & 0.68 & 2.28 & 97.72 & 0.74 & 0.39 & 51.56 & 48.44\\ \hline
644 & 12.31 & 1.39 & 11.32 & 88.68 & 1.43 & 0.62 & 43.34 & 56.66\\ \hline
649 & 68.54 & 26.18 & 38.20 & 61.80 & 36.14 & 33.39 & 92.40 & 7.60\\ \hline
654 & 96.51 & 40.26 & 41.71 & 58.29 & 48.22 & 17.09 & 35.43 & 64.57\\ 
\hline
\end{tabular}
\end{table*}

\begin{table*}
\centering
\caption{\spec integer rate cache results.}
\label{tab:specintrcache}
\begin{tabular}{|l|r|r|r|r|r|r|r|r|}\hline
\textbf{BM} & \textbf{L2 Access} & \textbf{L2 Miss} & \textbf{L2 Miss Rate} & \textbf{L2 Hit Rate} & \textbf{L3 Access} & \textbf{L3 Miss} & \textbf{L3 Miss Rate} & \textbf{L3 Hit Rate}\\ \hline
500 & 16.65 & 2.95 & 17.71 & 82.29 & 2.94 & 0.84 & 28.71 & 71.29\\ \hline
502 & 56.85 & 14.89 & 26.20 & 73.80 & 22.58 & 2.91 & 12.87 & 87.13\\ \hline
505 & 102.96 & 49.51 & 48.09 & 51.91 & 63.81 & 27.02 & 42.36 & 57.64\\ \hline
520 & 56.62 & 32.43 & 57.29 & 42.71 & 33.77 & 15.28 & 45.24 & 54.76\\ \hline
523 & 112.83 & 27.32 & 24.21 & 75.79 & 30.24 & 5.01 & 16.56 & 83.44\\ \hline
525 & 10.79 & 2.05 & 19.03 & 80.97 & 2.05 & 1.12 & 54.31 & 45.69\\ \hline
531 & 12.02 & 1.38 & 11.49 & 88.51 & 1.32 & 1.03 & 77.97 & 22.03\\ \hline
541 & 10.65 & 1.17 & 10.94 & 89.06 & 1.14 & 0.12 & 10.17 & 89.83\\ \hline
548 & 0.35 & 0.01 & 1.85 & 98.15 & 0.01 & 0.00 & 15.09 & 84.91\\ \hline
557 & 20.37 & 12.07 & 59.27 & 40.73 & 12.66 & 6.89 & 54.42 & 45.58\\ 
\hline
\end{tabular}
\end{table*}

\begin{table*}
\centering
\caption{\spec integer speed cache results.}
\label{tab:specintscache}
\begin{tabular}{|l|r|r|r|r|r|r|r|r|}\hline
\textbf{BM} & \textbf{L2 Access} & \textbf{L2 Miss} & \textbf{L2 Miss Rate} & \textbf{L2 Hit Rate} & \textbf{L3 Access} & \textbf{L3 Miss} & \textbf{L3 Miss Rate} & \textbf{L3 Hit Rate}\\ \hline
600 & 16.62 & 2.93 & 17.63 & 82.37 & 3.08 & 0.15 & 4.90 & 95.10\\ \hline
602 & 76.93 & 24.68 & 32.08 & 67.92 & 27.58 & 14.46 & 52.44 & 47.56\\ \hline
605 & 192.49 & 99.23 & 51.55 & 48.45 & 113.35 & 31.34 & 27.65 & 72.35\\ \hline
620 & 86.43 & 47.64 & 55.12 & 44.88 & 47.51 & 17.98 & 37.84 & 62.16\\ \hline
623 & 114.33 & 35.15 & 30.74 & 69.26 & 35.78 & 1.92 & 5.36 & 94.64\\ \hline
625 & 10.86 & 1.96 & 18.05 & 81.95 & 2.19 & 0.92 & 41.94 & 58.06\\ \hline
631 & 12.72 & 1.85 & 14.54 & 85.46 & 2.66 & 2.11 & 79.26 & 20.74\\ \hline
641 & 7.98 & 0.83 & 10.40 & 89.60 & 0.93 & 0.02 & 2.18 & 97.82\\ \hline
648 & 0.33 & 0.00 & 0.00 & 100.00 & 0.03 & 0.00 & 7.95 & 92.05\\ \hline
657 & 15.83 & 7.08 & 44.75 & 55.25 & 7.07 & 1.89 & 26.25 & 73.75\\  
\hline
\end{tabular}
\end{table*}

\begin{figure*}
    \centering
    \includegraphics[width=.94\textwidth]{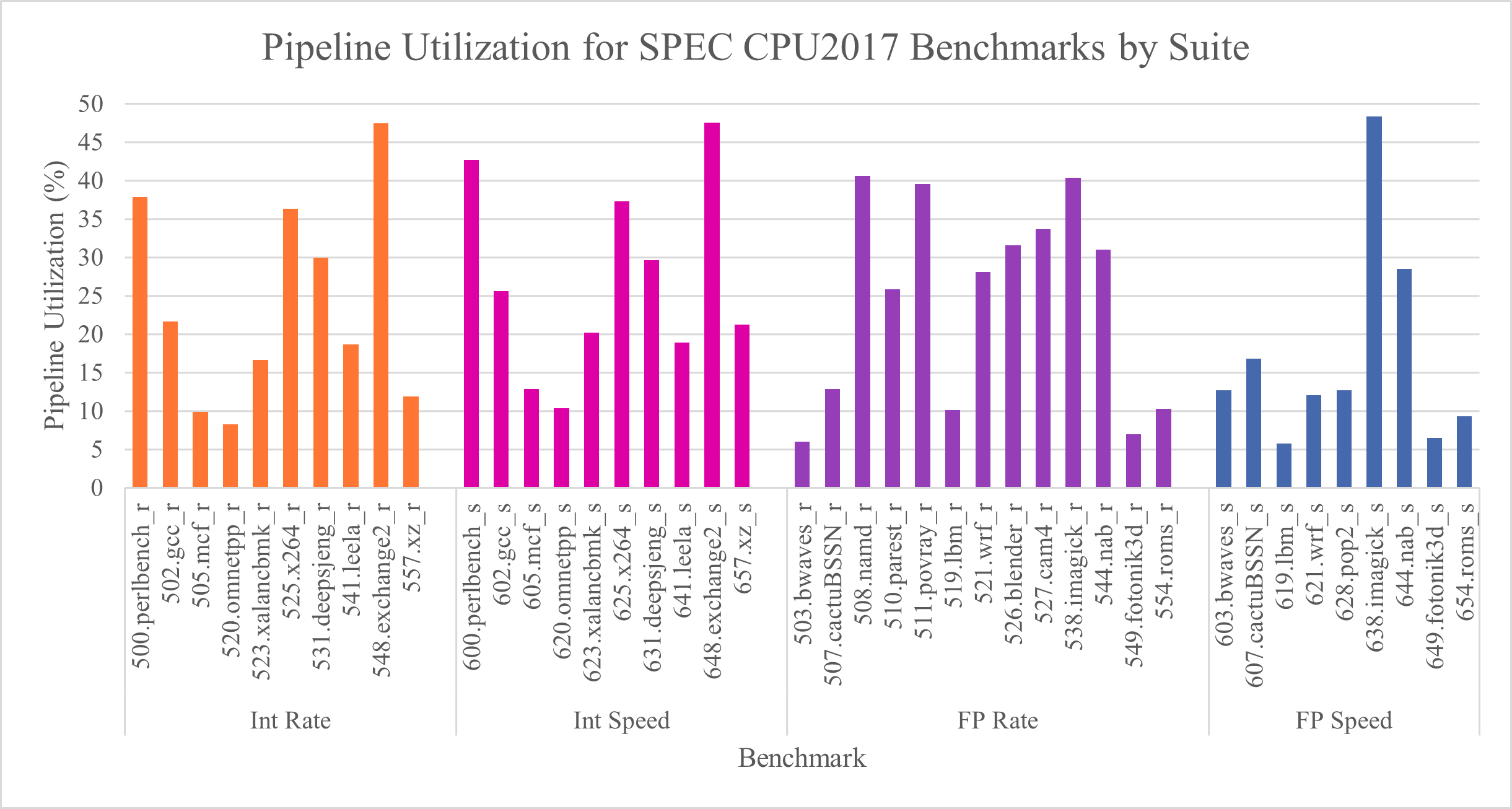}
    \caption{Pipeline utilization for all benchmarks.}
    \label{fig:pipe}
\end{figure*}

Tables~\ref{tab:specfprcache},~\ref{tab:specfpscache},~\ref{tab:specintrcache}, and~\ref{tab:specintscache} show the cache results from all four benchmark suites. All cache results sans miss and hit rates are shown as events per thousand instructions (PTI).
Pipeline utilization for all benchmarks can be seen in figure~\ref{fig:pipe}. The Rome architecture is capable of dispatching up to 6 macro-operations per clock cycle, thus pipeline utilization percent is calculated as $\textrm{IPC}/6\cdot 100$. Because this is a constant factor of IPC, this metric is merely an alternate visualization of IPC in a percentage.

\begin{figure*}
    \centering
    \includegraphics[width=.95\textwidth]{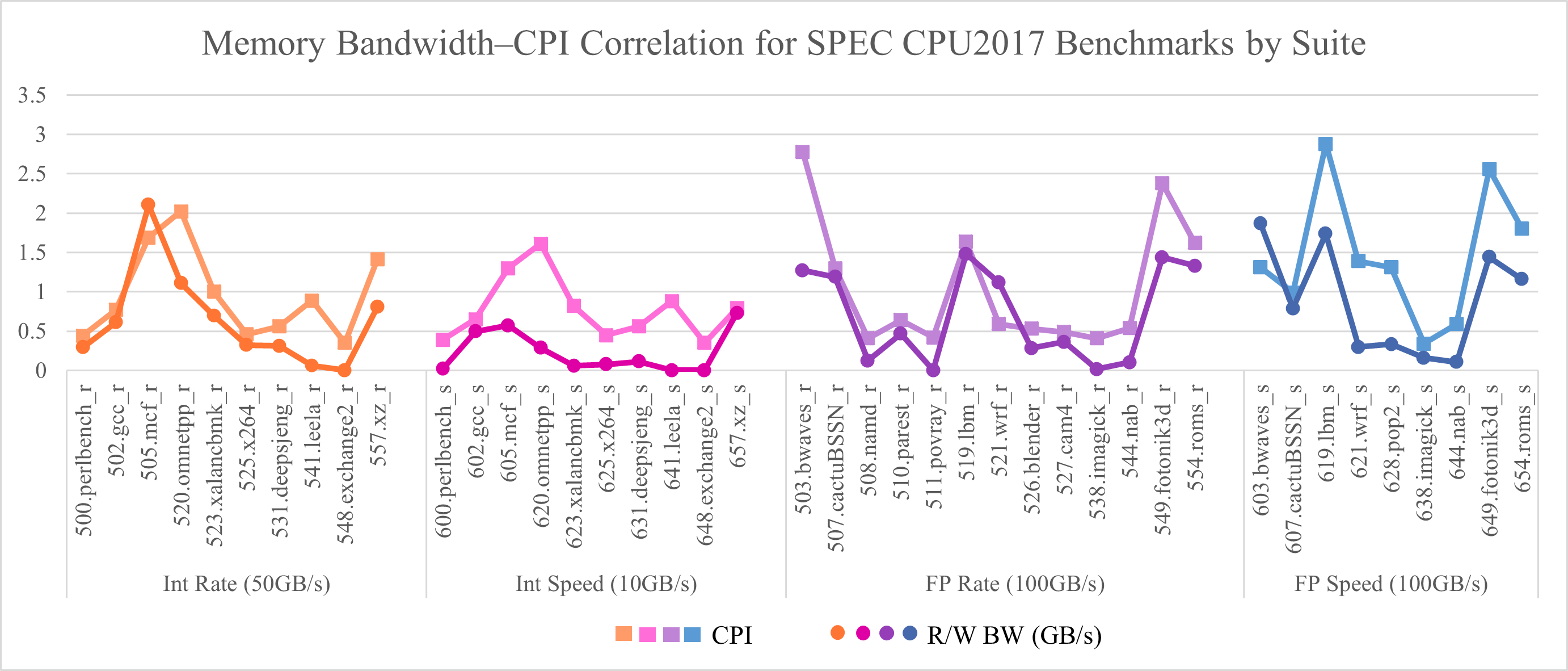}
    \caption{Memory bandwidth-CPI correlation for all benchmarks.}
    \label{fig:memcpi}
\end{figure*}

Figure~\ref{fig:memcpi} shows the correlation between CPI and memory bandwidth. There is a strong trend of higher memory bandwidth resulting in a higher CPI (lower IPC) across all benchmark suites. With data supported from all four benchmark suites, high-performance and compiler engineers should optimize code and code generation to reference main memory sparingly, as even slight increases in memory bandwidth can drastically raise CPI. Although it is a major engineering challenge, it is imperative for applications run on AMD Rome to lower reliance on the memory subsystem.

\begin{figure*}
    \centering
    \includegraphics[width=.95\textwidth]{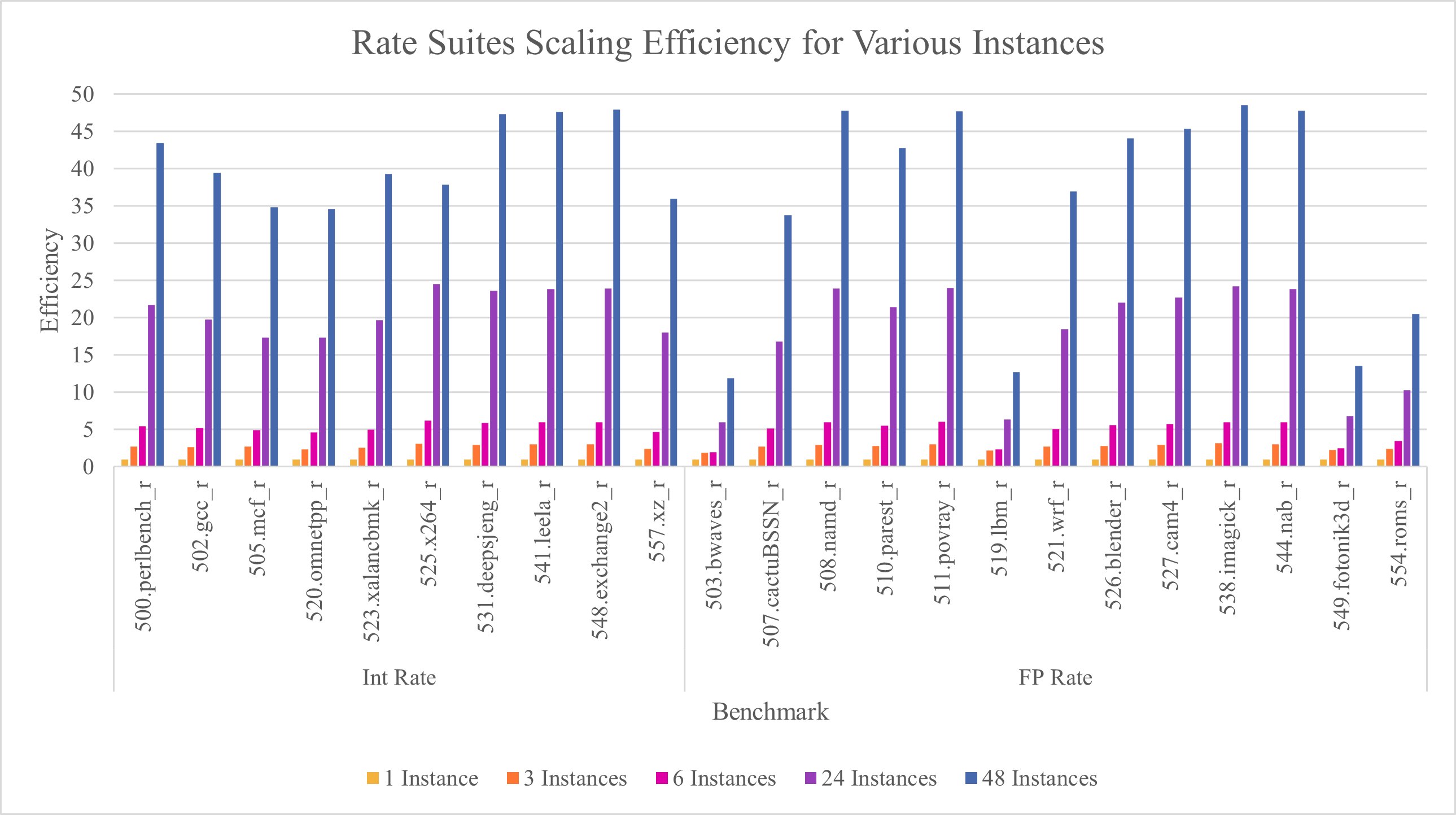}
    \caption{Metric $A$ for all rate suites.}
    \label{fig:rspeedup}
\end{figure*}

The overall pipeline utilization is unremarkable on the Epyc 7402. It is suspected that the NUMA cache requires many clock cycles to retrieve data. However, because the Rome architecture does not have counters for L3 hit latency (or if it does, aforementioned counters are undocumented), it is very difficult to analyze L3 hit latency of these benchmarks. However, from~\cite{romecache}, we know remote L3 accecss latencies for the Rome architecture are substantial, so any benchmark that regularly accesses an L3 cache owned by a remote CCX will suffer a $200+$ cycle penalty for each access.

However, a very clear trend emerges. Once the memory bandwidth surpasses about 37.5 GiB/s for integer benchmarks, or 100 GiB/s for floating-point benchmarks, the CPI increases considerably. This demonstrates the memory subsystem struggles to keep pace with requests from the CPU, signaling inefficiencies in the Rome architecture. In other words, the benchmark becomes memory-bottlenecked, which is indicative of a need for faster memory. Despite the memory subsystem on Euler being capable of 400 GiB/s, past this ``memory barrier'' the memory subsystem is unable to keep the CPU fed with data at a sufficiently fast rate.

This memory barrier corresponds to about 2.34 GiB/s per channel of memory for integer benchmarks, or 6.25 GiB/s for floating-point benchmarks.

\subsection{Scalability Test}

For the scalability test, as aforementioned we ran each rate benchmark with $i$ instances where $i\in\{1,3,6,24,48\}$. We used two metrics for measuring speedup demonstrated in two different charts. The first metric (metric $A$) is calculated with the following formula:

\[ A(bench, inst) = \frac{n\cdot\textrm{Time}(bench, 1)}{\textrm{Time}(bench, inst)} \]
where $\textrm{Time}(bench, inst)$ represents the arithmetic mean time it takes for $inst$ instances of $bench$ benchmark to complete. The second metric (metric $B$) is calculated as a geometric mean of the speedup across all rate benchmarks:

\[ B(inst) = \left(\prod_{b\in benches}\frac{\textrm{Time}(b, 1)}{\textrm{Time}(b, inst)}\cdot 100\right)^{1/|benches|} \]
where $\textrm{Time}(bench, inst)$ represents the arithmetic mean time it takes for $inst$ instances of $bench$ benchmark to complete, and $benches$ represents the set of all benchmarks in a suite.

Metric $A$ for all rate suites is shown in figure~\ref{fig:rspeedup}. Many benchmarks were able to approach a near-perfect speedup, though many, (especially in the floating-point rate suite) struggled to scale as contention for shared system resources rose. This is especially evident in floating point rate benchmarks 503.bwaves\_r, 519.lbm\_r, 549.fonotik3d\_r, and 554.roms\_r. These particular benchmarks were the four most demanding benchmarks on the memory subsystem according to figure~\ref{fig:memcpi}, consistent with the findings in~\cite{specmem}~\cite{ranjan}, and as a result had the highest CPI in the suite. Thus, it is evident these benchmarks are stressing the memory subsystem beyond CPI-efficient levels, and the system struggles substantially.

A similar trend arises with the integer rate benchmarks, as the 3 worst-scaling benchmarks were 505.mcf\_r, 520.omnetpp\_r, and 557.xz\_r. These 3 benchmarks also had the highest memory requirements and CPI according to figure~\ref{fig:memcpi}, though the effect is much less severe as opposed to the floating-point benchmarks. However, this is only due to the fact that the integer benchmarks were inherently less demanding on the memory subsystem as opposed to the floating-point benchmarks, also demonstrated in~\cite{specmem}.

\begin{figure}[t]
    \centering
    \includegraphics[width=.45\textwidth]{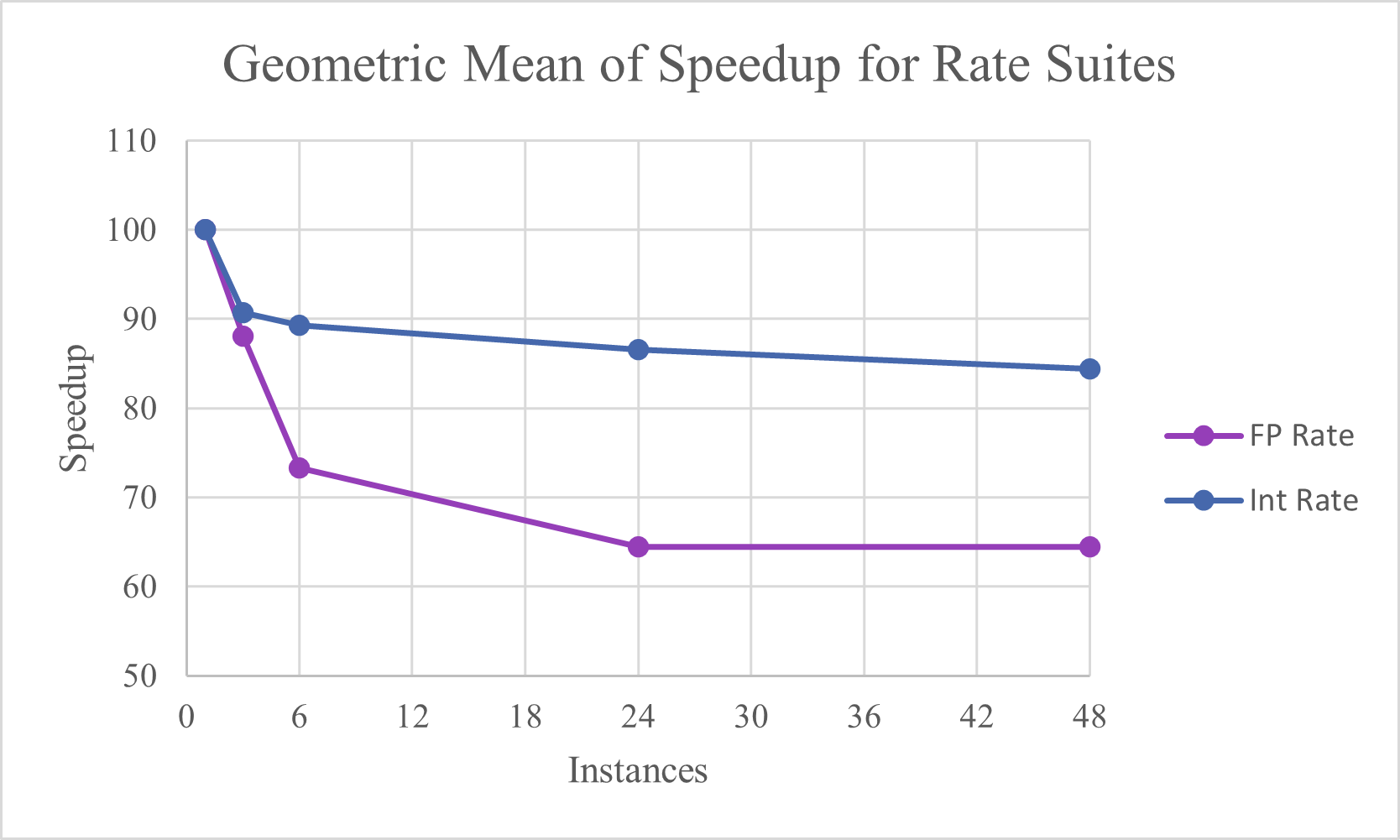}
    \caption{Metric $B$ for all rate suites.}
    \label{fig:rgeomean}
\end{figure}

Metric $B$ is shown in figure~\ref{fig:rgeomean}. These show the geometric mean of speedup across the entire suite. Because the floating-point benchmarks were more demanding on memory than the integer benchmarks, the floating-point benchmarks scale significantly worse. At 48 instances, the floating-point benchmarks have a speedup of only 64\%, while the integer benchmarks have a speedup of almost 85\%.

\section{Conclusions} \label{conclusions}
As novel architectures are released for public use, it is imperative that performance engineers analyze characteristics of said architectures to determine optimization techniques and strategies for leveraging aforementioned architecture to its fullest extent. This paper presents preliminary findings on the AMD Rome ``memory barrier,'' the maximum memory bandwidth which can be sustained without a significant impact on application performance. A clear wall representing efficient CPI is demonstrated when benchmarks from the \spec benchmark suite exceed 37.5 GiB/s memory bandwidth on integer benchmarks, or 100 GiB/s on floating-point benchmarks. To perform an adequate comparison, similar testing would need to be performed on a similarly-configured Intel system. With that information, high-performance engineers should have good characteristic information indicating which platform would be ideal for certain workloads. With this work, we hope compiler engineers and researchers can further leverage the AMD Rome architecture and have improved expectations for performance based on application characteristics.

\section*{Acknowledgement}
This work was made possible in part by hardware and software resources and technical support provided by the University of Alabama in Huntsville's Department of Electrical and Computer Engineering.

\bibliographystyle{IEEEtran}
\bibliography{conference}
\end{document}